\documentclass[letterpaper,12pt,twoside,final]{article}
\usepackage{showkeys}
\usepackage{natbib}
\usepackage{amsmath}
\usepackage{amsthm}
\usepackage{amscd}
\usepackage{amsfonts,amssymb}
\usepackage{url}
\def\urlprefix\url#1{\url{#1}} 

\newtheorem{theorem}{Theorem}

\newtheorem{lemma}{Lemma}

\theoremstyle{definition}

\theoremstyle{remark}
\newtheorem{remark}{Remark}

\newcommand{\N}{{\mathbb{N}}}
\newcommand{\R}{{\mathbb{R}}}
\newcommand{\Z}{{\mathbb{Z}}}
\newcommand{\FD}{{\mathcal{F}}}

\DeclareMathOperator{\ran}{ran}
\DeclareMathOperator{\const}{const}

\def\dd#1/#2.{{\frac{{}_{\displaystyle d #1}}{{}^{\displaystyle d #2}}}}
\def\pd#1/#2.{{\frac{{}_{\displaystyle \partial #1}}{{}^{\displaystyle \partial #2}}}}
\newcommand{\nfrac}[2]{{\frac{{}_{\displaystyle #1}}{{}^{\displaystyle #2}}}} 
\newenvironment{plainlist}{\list{}{\leftmargin0pt}}{\endlist}

\begin{document}
\title{Positive Measure Spectrum for Schr\"odinger Operators with Periodic Magnetic Fields }
\author{Michael J.\ Gruber\footnote{Department of Mathematics, The University of Arizona, Tucson AZ, USA}}


\maketitle
\begin{abstract}
We study Schr\"odinger operators with periodic magnetic field in $\R^2$, 
in the case of irrational magnetic flux.
Positive measure Cantor spectrum is generically expected in the presence of an electric potential.
We show that, even without electric potential, the spectrum has positive measure if the magnetic field is a perturbation of a constant one.

\end{abstract}

\section*{Introduction}
Magnetic Schr\"odinger operators have been studied in Solid State Physics, especially in connection with the Quantum Hall effect, as well as on their own right.
In a regular crystal `physics' is periodic, i.e., the electric potential - caused by the background field of the ions - is a periodic function. 
Magnetic fields - internal as well as external ones - are periodic as well, the latter ones typically being constant.
Alas, as is well known, magnetic fields enter the Schr\"odinger operator through a vector potential, so that the resulting operator is not necessarily periodic. Indeed, it is so only in the simple and well-understood case of `zero flux', where one has absolutely continuous spectrum and band-structure \citep{BirSus:TDPMHAC,Sob:ACPMSO}. Here, the magnetic flux (in units of flux quanta) is defined by
\begin{equation}
\Phi = \frac1{2\pi}\int_{\FD} B(x,y)\,dx\,dy,
\end{equation}
where $B$ is the magnetic field and $\FD$ is a lattice cell (fundamental domain for the action of the group). Note that in our units with $\hbar=e=1$, the magnetic flux quantum is just $2\pi$.

For integer or rational flux the spectrum will still consist of bands (possibly degenerating into points), but pure point spectrum is possible.

For irrational flux one expects Cantor spectrum (i.e.\ a nowhere dense set, no isolated points). 
The question is now: 
If $B$ is constant, what `defines the lattice', thus defining $\FD$ and the flux? 
A potential $V(x,y)=\cos (2\pi x) + \cos (2\pi y)$ has periods $(1,1)$; 
one might consider any pair $(m,n)$ of integers as periods of $V$, i.e., one can consider any coarser superlattice of $\Z\times\Z$ as the lattice of symmetry, but no finer lattice. 
Indeed, in this case one finds Cantor structure for a certain set of irrational values of $\Phi$ \citep{HelSjo:SAHEIIICSS}.

On the other hand, $V(x,y)=\cos(2\pi x)$ has periods $(1,c)$ for any real $c$; 
it does not define a fixed `minimally coarse' lattice. 
Indeed, the Schr\"odinger operator with constant magnetic field and this potential $V$ has band spectrum. 
This is still true for every potential $V(x,y)=V_0(x)$ with reasonably non-degenerate $V_0$.
Now, if we perturb such a $V$ by a periodic $V_1$ we expect Cantor spectrum, although this is not known.
We only know that the Sch\"odinger operator can be approximated in norm resolvent sense by operators from a natural algebra which have Cantor spectra \citep{Gru:NBT}.

Setting the Cantor issue aside, \cite{DinSinSos:SLLLSPLMSPP} showed that what survives under this kind of perturbation is the positive measure of the bands, although the bands might dissolve into a Cantor set.

We ask the same question for Schr\"odinger operators with periodic magnetic field $B$, without electric potential $V$.
For constant $B$ the spectrum is pure-point and infinitely degenerate (Landau levels).
Is a periodic (zero-flux) perturbation of $B$ enough to cause the same effects as the potential $V$?

In the course of giving an affirmative answer, we construct the generalized eigenfunctions and give estimates on the measure of the `bands'.

\subsection*{Outline}
In Section~\ref{sec:setup} we describe the setup and perform a first perturbation by one-dimensional magnetic potentials.
For the general case, we give the direct integral decomposition and
  express the operator in an appropriate basis `moving along the fiber'. 
In Section~\ref{sec:results} we give concise formulations of the main results.
The resulting double-infinite matrix problem is reduced in Section~\ref{sec:reduction} to a problem which is almost diagonal in a  sense made precise there.
In Section~\ref{sec:estimates} we prove the main reduction lemma, using estimates on Weber-Hermite functions.
Section~\ref{sec:proof} finishes the proof of the main results.

\subsection*{Acknowledgements}
We thank Leonid Friedlander for fruitful discussions and exchange of ideas.

\section{Setup and first perturbation} \label{sec:setup}
For simplicity of the notation we assume that the lattice is $\Z\times\Z$. 
Let $B$ be an arbitrary smooth periodic magnetic field and $\Phi$ its flux through a fundamental cell of the lattice
in units of flux quanta.
Then we can decompose $B$ as 
\begin{align*}
B&=B_c +B_z\quad\text{with} \\
B_c&= 2\pi \Phi\quad\text{and} \\
B_z&= B-B_c
\end{align*}
Note that $B_z$ has zero flux! 
Therefore, $B_z$ has a periodic vector potential $A_z$, and we can choose it to be of the form
\begin{align}
 A_z(x,y) &= \begin{pmatrix} \varepsilon_0 A^0(y) \\ \varepsilon_1 A^1(x,y) \end{pmatrix}
\end{align}
with periodic smooth $A^0$ and $A^1$, where we introduced parameters $\varepsilon_0,\varepsilon_1$ for later convenience.

$B_c$ is obviously constant, and we  choose a vector potential 
\begin{align}
A_z(y) &= B_c \begin{pmatrix} y \\ 0 \end{pmatrix}
\end{align}
for it.
In this gauge, the magnetic Schr\"odinger operator takes on the form
\begin{align}
H &= \frac12 \left[ \left( \nfrac1\imath \pd/x. - B_c y -\varepsilon_0 A^0(y) \right)^2 + \left( \nfrac1\imath\pd/y.-\varepsilon_1 A^1(x,y) \right)^2   \right]. \label{eq:operator}
\end{align}
Note that, in order to emphasize the dependencies, we write out the arguments of $A^0$ and $A^1$ even though they are multiplication operators; in the following we will keep sacrificing strictness of notation for better clarity in cases such as this one.

First we will set $\varepsilon_1=0$ and study the effect of switching on $\varepsilon_0$.
The operator~\eqref{eq:operator} does not depend on $x$ if $\varepsilon_1=0$.
Using Fourier transform on $L^2(\R_x)$, we can decompose $L^2(\R^2)$ into a direct integral $\int_\R^\oplus L^2(\R_y)\,d\xi$ such that 
\begin{subequations}
\begin{align}
L^2(\R^2)\ni f \mapsto \hat f,\quad 
  \hat f_\xi(y)&=\int_\R f(x,y)e^{-2\pi\imath \xi x}\,dx,   \\
  f(x,y) &= \int_\R f_\xi(y)e^{2\pi\imath \xi x}\,d\xi; \\
H = \int_\R^\oplus \hat H_\xi \, d\xi, \quad
  \hat H_\xi &= V_\xi(y)-\tfrac12\dd ^2/y^2. \label{eq:operatorfx} \quad\text{with} \\
V_\xi(y) &= \tfrac12  \left( 2\pi\xi - B_c y -\varepsilon_0 A^0(y) \right)^2. \label{eq:potentialfx}
\end{align}
\end{subequations}

In the case $\varepsilon_0=\varepsilon_1=0$ we are dealing with the Landau Hamiltonian.
We will go through its analysis since we will use its eigenfunctions later on as a basis.

\subsection{$\varepsilon_0=0=\varepsilon_1$} \label{case00}
\begin{align}
V_\xi(y) &= \tfrac12  \left( 2\pi\xi - B_c y \right)^2 = \tfrac{B_c^2}2 \left( y -  {\beta\xi} \right)^2
\end{align}
is a harmonic oscillator potential with frequency $B_c$, shifted by  $\beta\xi$ with $\beta=\frac{2\pi}{B_c}=\frac1{\Phi}$ (we assume $B_c\ne0$, or else there is not much to do).
If we denote by 
\begin{align}
\Omega_m(y)=\frac{(-1)^m}{\sqrt{\sqrt \pi 2^mm!}}\exp\left(\tfrac{y^2}2\right) \dd^m/y^m. \exp\left(-y^2\right),\quad m\in\Z_+,
\end{align}
the Weber-Hermite functions, i.e.\ the standard normalized eigenfunctions of the harmonic oscillator with frequency $1$,
then
\begin{align}
\Psi_{\xi,m} &= \sqrt[4]{B_c}\, \Omega_m\left( \sqrt{B_c}\left(y- {\beta\xi}  \right)\right),\quad m\in\Z_+, \label{eq:eigenfunctions}
\end{align}
are the eigenfunctions of $\hat H_\xi$, and the corresponding eigenvalues are $B_c\left(\frac12+m\right)$.
Since the spectrum of $\hat H_\xi$ is independent of $\xi$ it coincides with the spectrum of $H$ as a set, and both are pure-point.
On the other hand, since $H$ is invariant in $x$, it has infinitely degenerate spectrum.

\subsection{$\varepsilon_0\ne0=\varepsilon_1$}
Introducing the shifted variable $\tilde y=y-{\beta\xi}$, we can write the potential~\eqref{eq:potentialfx} as
\begin{align}
V_\xi(\tilde y) &= \tfrac{B_c^2}2 \left( \tilde y + \varepsilon_0 A^0\left(\tilde y+{\beta\xi}\right)\right)^2.
\end{align}
Since $A^0$ is periodic and smooth (and therefore bounded), this is just a perturbation of the harmonic oscillator potential.
$V_\xi(\tilde y)$ will still tend to infinity as $|\tilde y|$ does, so that $\hat H_\xi$ has discrete spectrum.
Some simple estimates using test functions show that for small $\varepsilon_0$ the eigenvalues will be within $C_m\varepsilon_0\max|A^0|$ of the Landau levels, for some constants $C_m$ (involving the maximum of $|y\Omega_m(y)|$) depending on $m$ and $B_c$ only; in particular, they are independent of $\xi$.
A closer investigation shows that $C_m$ is indeed bounded with respect to $m$.

Note that $V_\xi$ is periodic in $\xi$ with period $\frac{B_c}{2\pi}=\Phi$.
Therefore, the spectrum of $H=\int_\R^\oplus \hat H_\xi\,d\xi$ consists of bands whose size is bounded by $2C_m\varepsilon_0\max|A^0|$.
These bands might degenerate into points.

\subsection{$\varepsilon_0\ne0\ne\varepsilon_1$}
Now $H$ is not independent of $x$ any more.
But at least it will be periodic in $x$ with period $1$, since $A^1$ is periodic.
Using Fourier series on $L^2([0,1]_x)$, we can decompose  $L^2(\R^2)$ into a direct integral $\int_{[0,1]}^\oplus L^2([0,1]_x\times\R_y)\,d\xi$ such that
\begin{subequations}
\begin{align}
L^2(\R^2)\ni f \mapsto \hat f,\quad 
  \hat f_\xi(x,y)&=\sum_{l\in\Z} f(x+l,y)e^{-2\pi\imath \xi(x+l)} ,  \\
  f(x,y) &= \int_{[0,1]} \hat f_\xi(x,y)e^{2\pi\imath\xi x}\,d\xi; \\
H = \int_{[0,1]}^\oplus \hat H_\xi \, d\xi, \quad
  \hat H_\xi &=  \frac12 \left[ \left( \nfrac1\imath \pd/x. +2\pi\xi - B_c y -\varepsilon_0 A^0(y) \right)^2 +\right. \notag \\
  &\quad + \left. \left( \nfrac1\imath\pd/y.-\varepsilon_1 A^1(x,y) \right)^2   \right] 
\end{align}
\end{subequations}
acting on functions periodic in $x$. 
Note that we keep denoting the fibre operator $\hat H_\xi$ for the new direct integral, in order to avoid an inflation of notation.

If we now choose the basis $\left(e^{2\pi\imath nx}\right)_{n\in\Z}$ in $L^2([0,1]_x)$, defining an isomorphism with $\ell^2(\Z)$, and combine this with the isomorphism defining the direct integral above, we arrive at the direct integral $L^2(\R^2)=\int_{[0,1]}^\oplus \ell^2(\Z)\otimes L^2(\R)\,d\xi$ with
\begin{subequations}
\begin{align}
L^2(\R^2)\ni f \mapsto \hat f,\quad 
  \hat f_{\xi,n}(y)&=\int_{[0,1]} \sum_{l\in\Z} f(x+l,y)e^{-2\pi\imath \left[ \xi l+(\xi+n)x \right] } \,dx,  \\
  f(x,y) &= \int_{[0,1]} \sum_{n\in\Z} \hat f_{\xi,n}(y)e^{2\pi\imath(\xi+n) x}\,d\xi; \\
H = \int_{[0,1]}^\oplus \hat H_\xi \, d\xi, \quad
  \hat H_\xi &=  \frac12 \left[ \left( 2\pi(\nu+\xi) - B_c y -\varepsilon_0 A^0(y) \right)^2 +\right. \notag \\
  &\quad + \left. \left( \nfrac1\imath\dd/y.-\varepsilon_1 \widehat{ A^1}(y)\star \right)^2   \right] \label{eq:preprop}
\end{align}
\end{subequations}
Here, $\nu$ is the operator of multiplication on $\ell^2(\Z)$, i.e. $(\nu g)(n)=ng(n)$, and $\widehat{ A^1}(y)\star$ is convolution with the Fourier series of $A^1(x,y)$ in $x$:
\begin{subequations}
\begin{align}
(\widehat{ A^1}(y)\star g)(n) &= \sum_{l\in\Z} g(l)\widehat{ A^1}_{n-l}(y), \\
 \widehat{ A^1}_n(y) &= \int_{[0,1]}A^1(x,y)e^{-2\pi\imath nx}\,dx
\end{align}
\end{subequations}
Note that, of course, our basis functions belong to the domain of the operator.

As a final step, we choose a special basis in $L^2(\R)$, namely the eigenfunctions $\left(\Psi_{\xi,m}\right)_{m\in\Z_+}$ described in subsection~\ref{case00}, equation~\eqref{eq:eigenfunctions}.
Thus we arrive at
\begin{lemma} \label{lemma:direct integral}
There is a decomposition of $L^2(\R^2)$ into $\int_{[0,1]}^\oplus \ell^2(\Z\times \Z_+)\,d\xi$ with
\begin{subequations}
\begin{align}
L^2(\R^2)\ni f &\mapsto \hat f,\\
  \hat f_{\xi,n,m}&=\int_{[0,1]\times\R} \sum_{l\in\Z} f(x+l,y)e^{-2\pi\imath \left[ \xi l+(\xi+n)x \right] }\Psi_{\xi+n,m}(y) \,dx\,dy,  \\
  f(x,y) &= \int_{[0,1]} \sum_{n\in\Z} \sum_{m\in\Z_+}\hat f_{\xi,n,m}e^{2\pi\imath(\xi+n) x}\Psi_{\xi+n,m}(y)\,d\xi 
\end{align}
\end{subequations}
such that $H$ decomposes as
\begin{subequations}
\begin{align}
 H &= \int_{[0,1]}^\oplus \hat
 H_\xi \, d\xi, \\
  \left(\hat H_\xi \hat f_{\xi}\right)_{m,n} &=  B_c\left(\tfrac12+m\right)\hat f_{\xi,m,n} \label{eq:I}\\
 &\quad + \frac12 \varepsilon_0^2 \sum_{l\in\Z_+} \widetilde{ A^0}_{l,m}\left({\beta(\xi+n)} \right)\hat f_{\xi,l,n}\label{eq:II} \\ 
 &\quad  - \varepsilon_0 \sum_{l\in\Z_+} A^0_{l,m} \left({\beta(\xi+n)} \right) \hat f_{\xi,l,n} \label{eq:III}\\
 &\quad +\frac12\varepsilon_1^2 \sum_{l\in\Z_+}\sum_{k\in\Z}  \widetilde{ A^{1}}^{(n-k)}_{l,m}\left({\beta(\xi+n)} \right)   \hat f_{\xi,l,k} \label{eq:IV} \\
 &\quad - \varepsilon_1 \sum_{l\in\Z_+}\sum_{k\in\Z} \widehat{ A^{1}}^{(n-k)}_{l,m}\left({\beta(\xi+n)} \right)   \hat f_{\xi,l,k} \label{eq:V}  
\end{align}
\end{subequations}
where
\begin{subequations}
\begin{align}
 \widetilde{ A^0}_{l,m}(p) &= \sqrt{B_c} \int_\R (A^0)^2(y+p)\Omega_l\left(\sqrt{B_c}y\right)\Omega_m\left(\sqrt{B_c}y\right)\,dy, \label{eq:II=}\\ 
 A^0_{l,m} (p) &= {B_c}^{\frac32} \int_\R A^0(y+p)y\Omega_l\left(\sqrt{B_c}y\right)\Omega_m\left(\sqrt{B_c}y\right)\,dy, \label{eq:III=}\\
 \widetilde{ A^{1}}^{(k)}_{l,m}(p) &= \sqrt{B_c} \int_\R \widehat{(A^1)^2}_{k}(y+p)\Omega_l\left(\sqrt{B_c}\left(y+{\beta k}\right)\right)\Omega_m\left(\sqrt{B_c}y\right)\,dy, \label{eq:IV=} \\
 \widehat{ A^{1}}^{(k)}_{l,m}(p) &= \nfrac1\imath B_c \int_\R \widehat{A^1}_{k}(y+p) \left[ \Omega_l'\left(\sqrt{B_c}\left(y+{\beta k}\right)\right)\Omega_m\left(\sqrt{B_c}y\right)\right. \notag \\
&\quad - \left . \Omega_l\left(\sqrt{B_c}\left(y+{\beta k}\right)\right)\Omega_m'\left(\sqrt{B_c}y\right)\right] \,dy. \label{eq:V=}
\end{align}
\end{subequations}
As above, $\widehat{A^1}_k$ denotes the $k$-th Fourier coefficient of $A^1$ with respect to $x$, and analogously for $(A^1)^2$.
\end{lemma}

\begin{proof}
We have 
\[ \left(\hat H_\xi \hat f_{\xi}\right)_{m,n} =  \sum_{l\in\Z_+}\sum_{k\in\Z} \hat H_{\xi,m,n;l,k} \hat f_{\xi,l,k} \]
so that we just have to compute the matrix elements in the given basis, for all the terms in~\eqref{eq:preprop}.
\begin{plainlist}
\item[\eqref{eq:I}] These are just the Landau eigenvalues in the case $\varepsilon_0=0=\varepsilon_1$.
\item[\eqref{eq:II} \& \eqref{eq:II=}] For the terms with coefficient $\varepsilon_0^2$  we have to compute the matrix element of the square term $(A^0)^2$ which is
\begin{align*}
 & \sqrt{B_c} \int_\R (A^0)^2(y)\Omega_m\left(\sqrt{B_c}\left(y-{\beta(\xi+n)}\right)\right)\Omega_l\left(\sqrt{B_c}\left(y-{\beta(\xi+n)}\right)\right)\,dy 
\end{align*}
\item[\eqref{eq:III} \& \eqref{eq:III=}] The terms with coefficient $\varepsilon_0$ in~\eqref{eq:preprop} give the matrix element of the mixed term $(B_cy-2\pi(\nu+\xi))A^0$, the calculation is the same as above.
\item[\eqref{eq:IV} \& \eqref{eq:IV=}] The coefficient $\varepsilon_1^2$ singles out the square term $\widehat{ A^{1}}\star\widehat{ A^{1}}\star=\widehat{ (A^{1})^2}\star$, i.e.\ convolution with the Fourier series of ${ (A^{1})^2}$. 
Its matrix element is
\begin{align*}
 & \sqrt{B_c} \int_\R \widehat{(A^1)^2}_{n-k}(y)\Omega_m\left(\sqrt{B_c}\left(y-{\beta(\xi+n)}\right)\right)\Omega_l\left(\sqrt{B_c}\left(y-{\beta(\xi+k)}\right)\right)\,dy 
\end{align*}
so that shifting $y$ as above gives the desired result.
\item[\eqref{eq:V} \& \eqref{eq:V=}] The term with coefficient $\varepsilon_1$ is $\imath\left(\dd/y.\circ \widehat{ A^{1}}\star  + \widehat{ A^{1}}\star \dd/y.\right) $.
The first part can be written $\dd/y.\circ \widehat{ A^{1}}\star=\widehat{ A^{1}}' \star + \widehat{ A^{1}}\star\dd/y.$ which has matrix element
\begin{align*}
 & \sqrt{B_c} \int_\R \left[ \widehat{A^1}'_{n-k}(y)\Omega_m\left(\sqrt{B_c}\left(y-{\beta(\xi+n)}\right)\right)\Omega_l\left(\sqrt{B_c}\left(y-{\beta(\xi+k)}\right)\right) \right. \\
&\quad + \sqrt{B_c} \left. \widehat{A^1}_{n-k}(y)\Omega_m\left(\sqrt{B_c}\left(y-{\beta(\xi+n)}\right)\right)\Omega_l'\left(\sqrt{B_c}\left(y-{\beta(\xi+k)}\right)\right) \right] \,dy.
\end{align*}
On the other hand, we can use partial integration for the matrix element of the second part, which is $\widehat{ A^{1}}\star \dd/y.$:
\begin{align*}
&{B_c} \int_\R \widehat{A^1}_{n-k}(y)\Omega_m\left(\sqrt{B_c}\left(y-{\beta(\xi+n)}\right)\right)\Omega_l'\left(\sqrt{B_c}\left(y-{\beta(\xi+k)}\right)\right)  \,dy \\
 &=  \sqrt{B_c} \int_\R \left[ - \widehat{A^1}'_{n-k}(y)\Omega_m\left(\sqrt{B_c}\left(y-{\beta(\xi+n)}\right)\right)\Omega_l\left(\sqrt{B_c}\left(y-{\beta(\xi+k)}\right)\right) \right. \\
&\quad - \sqrt{B_c}\left. \widehat{A^1}_{n-k}(y)\Omega_m'\left(\sqrt{B_c}\left(y-{\beta(\xi+n)}\right)\right)\Omega_l\left(\sqrt{B_c}\left(y-{\beta(\xi+k)}\right)\right) \right] \,dy.
\end{align*}
These two parts add up to the desired result.
\end{plainlist}
\end{proof}

\begin{remark}
Note that the functions $\widetilde{ A^0}_{l,m}(p)$, $A^0_{l,m} (p) $, $\widetilde{ A^{1}}^{(k)}_{l,m}(p)$, $ \widehat{ A^{1}}^{(k)}_{l,m}(p)$ have period $1$ in $p$.
Also, due to the decay of the Weber-Hermite functions $\Omega_m(y)$ in $y$, the functions  $\widetilde{ A^{1}}^{(k)}_{l,m}(p)$, $ \widehat{ A^{1}}^{(k)}_{l,m}(p)$ exhibit exponential decay in $k$.
\end{remark}

\begin{remark} \label{remark:x only}
If $A^1$ depends on $x$ only then $\widetilde{ A^{1}}^{(k)}_{l,m}(p)=\delta_{k,0}\delta_{l,m}\overline{(A^1)^2}$ with the average $\overline{(A^1)^2}$ of $(A^1)^2$ with respect to $x$.
If we assume that $\overline{A^1}=0$ (we can always add a constant to $A^1$ to achieve this, without changing the magnetic field) then $ \widehat{ A^{1}}^{(k)}_{l,m}(p)=0$.
\end{remark}

\begin{remark} \label{remark:cos}
In the simplest non-trivial case $A^0(y)=\cos(2\pi y)$, using parity of the Weber-Hermite functions we get
\begin{align*}
{ A^0}_{l,m}(p) &= a_{l,m}\cos(2\pi p)+b_{l,m}\sin(2\pi p),
\end{align*}
where $a_{l,m}=0$ if $m+l$ is even, and $b_{l,m}=0$ if $m+l$ is odd. Similarly,
\begin{align*}
\widetilde{ A^0}_{l,m}(p) &= \tfrac12\delta_{l,m}+c_{l,m}\cos(4\pi p)+d_{l,m}\sin(4\pi p),
\end{align*}
where $c_{l,m}=0$ if $m+l$ is odd, and $d_{l,m}=0$ if $m+l$ is even.
\end{remark}

\begin{remark} \label{remark:weber hermite}
Creation and annihilation operators on harmonic oscillator functions yield the relations
\begin{subequations}
\begin{align}
\Omega_m'(y) &= \sqrt{\tfrac{m}2}\Omega_{m-1} - \sqrt{\tfrac{m+1}2}\Omega_{m+1}, \label{eq:Omega'} \\
y\Omega_m(y) &= \sqrt{\tfrac{m}2}\Omega_{m-1} + \sqrt{\tfrac{m+1}2}\Omega_{m+1}. \label{eq:yOmega} 
\end{align}
\end{subequations}
Using these we can express ${ A^0}_{l,m}$ and $ \widehat{ A^{1}}^{(k)}_{l,m}$ solely in terms of $A^0$, $A^1$ and Weber-Hermite functions, without referring to their derivatives or multiplication by $y$.
\end{remark}

\section{Main results} \label{sec:results}
Let $P_{\xi,m}^0$ denote the projection on the eigenspace of the $m$-th Landau level.
Lemma~\ref{lemma:direct integral} tells us that the action of $\hat H_\xi$ in this eigenspace, i.e.\ the
part of $\hat H_\xi$ which is `diagonal in $m$', amounts to:
\begin{align}
& \left( P_{\xi,m}^0 \hat H_\xi  P_{\xi,m}^0 \right) \hat f_{\xi,m,n} \notag \\ 
&= \left( B_c\left(\tfrac12+m\right)
 + \tfrac12 \varepsilon_0^2  \widetilde{ A^0}_{m,m}\left({\beta(\xi+n)} \right) 
  - \varepsilon_0  A^0_{m,m} \left({\beta(\xi+n)} \right)  \right) \hat f_{\xi,m,n} \label{eq:almost Matthieu} \\
 &\quad + \sum_{k\in\Z} \left[ \tfrac12\varepsilon_1^2 \widetilde{ A^{1}}^{(n-k)}_{m,m}\left({\beta(\xi+n)} \right)  
  - \varepsilon_1  \widehat{ A^{1}}^{(n-k)}_{m,m}\left({\beta(\xi+n)} \right) \right]  \hat f_{\xi,m,k} \notag 
\end{align}
For fixed $\xi$ and $m$ this is a one-dimensional difference operator with quasiperiodic coefficients and exponentially decaying off-diagonal (i.e.~$k\ne0$) terms. 
If we choose $A^1$ to be independent of $y$ as in Remark~\ref{remark:x only} there are no off-diagonal terms at all.
If we furthermore choose $A^0(y)=\cos(2\pi y)$ as in Remark~\ref{remark:cos} then \eqref{eq:almost Matthieu} will be similar to the Almost Matthieu operator, with a slightly more complicated potential.
Indeed, if we look at terms of order up to $\varepsilon_0$ only it will be exactly the Almost Matthieu operator.
In the case of constant magnetic field this observation goes back to \cite{Hof:ELWFBERIMF}.

If $\varepsilon_0$ and $\varepsilon_1$ are small enough then $H$ will have invariant subspaces $E_m$ which are close to the eigenspaces of the Landau levels.
We will construct a unitary transformation to achieve the following:
\begin{theorem} \label{theorem:direct integral}
For every small enough $\varepsilon_0$ and $\varepsilon_1$ there is an $M(\varepsilon_0, \varepsilon_1)$ such that $M(\varepsilon_0, \varepsilon_1)\to\infty$ as both $\varepsilon_0, \varepsilon_1\to\infty$, 
and such that for $m\leq M(\varepsilon_0,\varepsilon_1)$ the invariant subspace $E_m$ of H and the restriction of $H$ to $E_m$ have a direct integral decompostion
\begin{subequations}
\begin{align}
E_m &= \int_{[0,1]}^\oplus \ell^2(\Z)\,d\xi, \\
 H_m:=H\vert_{E_m} &= \int_{[0,1]}^\oplus H_{m,\xi}\,d\xi.
\end{align}
\end{subequations}
Furthermore, $H_{m,\xi}$ acts on $g\in\ell^2(\Z)$ as a one-dimensional difference operator with exponentially decaying coefficients close to those of~\eqref{eq:almost Matthieu}:
\begin{subequations}
\begin{gather}
 \left(H_{m,\xi}g\right)(n) 
= d_m\left({\beta(\xi+n)} \right) g(n)
 + \sum_{k\in\Z} a_m(n-k,{\beta(\xi+n)} )g(k), \\
 \left\| d_m(\cdot) -
 \left[ B_c\left(\tfrac12+m\right)
   - \varepsilon_0  A^0_{m,m} (\cdot)  \right] \right \|
_{C^2(S^1)}  < C_0 \varepsilon_0^2, \\
 \sum_{k\in\Z\setminus\{0\}} \left\| a_m(k,\cdot)e^{\delta|k|} \right \|
_{C^2(S^1)}  < C_1 \varepsilon_0
\end{gather}
\end{subequations}
for some $\delta >0$
\end{theorem}
We denote by $\|\cdot\|_{C^2(S^1)}$ the sum of the supremum norms of the derivatives of order $0,1,2$.

As in \citep{DinSinSos:SLLLSPLMSPP} we can now make use of the results of \cite{Din:SPSTDOQC} on ergodic families of operators.
In order to apply these results we need the following assumptions:
\begin{description}
\item[diophantine] There are $C>0$, $\kappa>0$ such that $|\{\beta n\}|>C/|n|^\kappa$ for all $n\in\Z\setminus{0}$.
Here, $\beta=\tfrac{2\pi}{B_c}=\tfrac1\Phi$ as before, and $\{\cdot\}$ denotes the fractional part.
\item[smoothness] $A^0$ and $A^1(x,y)$ are smooth; furthermore, all derivatives $\pd^j A^1/y^j.(x,y)$ are analytic in $|\Im x|<\delta$ for some common $\delta>0$.
\item[Morse] $A^0_{m,m}$ is a Morse function on $S^1$ with exactly two critical points.
\end{description}
\begin{theorem} \label{theorem:main}
Let $\beta$ be diophantine and $A^0$, $A^1$ smooth as defined above, and $M>0$ such that $A^0_{m,m}$ is a Morse function with exactly two critical points for all $m\leq M$.
Then there are $\tilde\varepsilon_0,\tilde\varepsilon_1>0$ depending on $M$ such that for all $\varepsilon_0<\tilde\varepsilon_0,\varepsilon_1<\tilde\varepsilon_1$ the following are true for all $m\leq M$:
\begin{enumerate}
\item \label{eigenvalues} There are $1$-periodic measurable functions $\lambda_m$ such that for every $n\in\Z$ and almost every $\xi\in[0,1]$, $\lambda_m(\beta(\xi+n))$ is an eigenvalue of $H_{m,\xi}$ and therefore in the spectrum of $H_m$. 
Furthermore,
\begin{align}
 \left\| \lambda_m(\cdot) -
 \left[ B_c\left(\tfrac12+m\right)
   - \varepsilon_0  A^0_{m,m} (\cdot)  \right] \right \|
_{L^\infty(S^1)}  < \const \varepsilon_0^2. \label{eq:eigenvalues}
\end{align}

\item \label{eigenfunctions} There are $1$-periodic measurable functions $f_{m,l,k},l\in\Z_+,n\in\Z,$ decaying exponentially in $n$ such that for almost every $\xi\in[0,1]$,
\begin{align}
\sum_{\substack{ l\in\Z_+\\k\in\Z\phantom{{}_+}} } |f_{m,l,n}(\beta\xi)|(l^2+1)e^{2\delta|n|}<\infty, \label{eq:maincoefficients}
\end{align}
and for every $k\in\Z$, the series
\begin{align}
\Phi_{m,\xi,k}(x,y) &= \sum_{\substack{ l\in\Z_+\\n\in\Z\phantom{{}_+}} } f_{m,l,n-k}(\beta(\xi+k))e^{2\pi\imath(\xi+n)x} \Psi_{\xi+n,l}(y)  \label{eq:maineigenfunctions}
\end{align}
and all its derivatives converge uniformly in $x,y$.
$\Phi_{m,\xi,k}$ is an eigenfunction of $H_{m,\xi}$ and therefore a generalized eigenfunction of $H$, with eigenvalue $\lambda_m(\beta(\xi+k))$. 
Moreover, for every $N>0$ and for $\varepsilon_0,\varepsilon_1$ small enough (depending on $N$),
\begin{align}
\left| \Phi_{m,\xi,k}(x,y)  \right| \leq \frac{\const}{y^{2N}+1} \label{eq:maindecay}
\end{align}
with the constant depending on $\xi,k,m,N$.
\item \label{spectrum} $H_{m}$ is uniformly $\varepsilon_0$-close to  band structure:  $H_{m}$ is unitarily equi\-va\-lent to multiplication by the function $\lambda_m(\beta\cdot)$.
The Lebesgue measure of the spectrum of $H_m$ is $\varepsilon_0|\ran A^0_{m,m}|+O(\varepsilon_0^2)$.

\end{enumerate}
\end{theorem}

\begin{remark}
One can reduce the smoothness requirements somewhat (to analyticity in $x$ of a finite number of derivatives in $y$), thereby weakening the result on decay of the generalized eigenfunctions in $y$.
\end{remark}

\begin{remark}
Of course one can also include an electric potential into the picture; 
\cite{DinSinSos:SLLLSPLMSPP} did so in the case of constant magnetic field, i.e.\ $A^0\equiv A^1\equiv0$.
The point in our work is that the magnetic field perturbation alone is strong enough to deform the Landau levels into a spectral set with positive measure.
\end{remark}

\section{Reduction} \label{sec:reduction}
$\hat H_\xi$ is a double matrix operator on $\ell^2(\Z_+\times\Z)$ with indices $(m,n)\in\Z_+\times\Z$.
We decompose it as $\hat H_\xi=\mathcal{D}_1+\mathcal{M}_1+\mathcal{O}_1$, where $\mathcal{D}_1$ is diagonal in $m$ and $n$, $\mathcal{O}_1$ is off-diagonal in $m$ and contains only the first row and the first column, and $\mathcal{M}_1$ is the remainder.
Note that both $\mathcal{O}_1$ and $\mathcal{M}_1$ are of order $\varepsilon_0$ (we always assume $\varepsilon_1<\varepsilon_0$).

Our goal is to find a unitary transformation $U$ which kills the terms in $\mathcal{O}$ (they represent the interaction between different Landau levels).
$U$ should  leave the rest of the structure basically untouched.
We will show how to accomplish this for the interaction between the $0$-th and all other bands; extending this to $M$ off-diagonal terms is a trivial generalisation.

The strategy is as follows: We define $U$ as $U=\prod_{j\in\N}U_j$, where $U_j$ eliminates off-diagonal terms up to (and including) order $\varepsilon_0^j$.
Each transformation is of the form $U_j=e^{\imath W_j}$ for a Hermitian bounded $W_j$ whose coefficients are of order $\varepsilon_0^j$.
We use the Baker-Hausdorff formula
\begin{subequations}
\begin{align}
e^{-\imath A}Be^{\imath A} &= \sum_{r\in\Z_+}\frac{\imath^r}{r!}[B,A]_r, \\
 [B,A]_0 &= B, \\
 [B,A]_{r+1} &= [ [B,A]_r,A]
\end{align}
\end{subequations}
in order to find $W_j$.
In fact, in the $j$-th step we will only have to consider the terms up to order $\varepsilon_0^j$ in this formula, which are the terms $r=0$ and $r=1$. 
This gives us the equation 
\[ \mathcal{O}_1 + \imath [\mathcal{D}_1,W_1] =0, \]
which reads as follows for the coefficients:
\begin{subequations}
\begin{gather}
W_1(m,n;l,k) = 0\quad\text{for}\quad m>0 ,\\
\mathcal{O}_1(0,n;l,k)+\imath\left[\mathcal{D}_1(0,n;0,n)-\mathcal{D}_1(l,k;l,k) \right] W_1(0,n;l,k)=0 \label{eq:small denominator}
\end{gather}
\end{subequations}
Because of
\begin{gather*}
\left|\mathcal{D}_1(0,n;0,n)-\mathcal{D}_1(l,k;l,k) \right| > \\ 
B_cl 
-\tfrac12\varepsilon_0^2\left( \left\|\widetilde{A^0}_{0,0}\right\|_\infty+ \left\|\widetilde{A^0}_{l,l}\right\|_\infty  \right) 
- \varepsilon_0\left( \left\|{A^0}_{0,0}\right\|_\infty+ \left\|{A^0}_{l,l}\right\|_\infty  \right) \\
- \tfrac12\varepsilon_1^2\left( \left\|\widetilde{A^1}_{0,0}^{(0)}\right\|_\infty+ \left\|\widetilde{A^1}_{l,l}^{(0)}\right\|_\infty  \right) 
- \varepsilon_1\left( \left\|\widehat{A^1}_{0,0}^{(0)}\right\|_\infty+ \left\|\widehat{A^1}_{l,l}^{(0)}\right\|_\infty  \right)\\
> \tfrac12 B_cl
\end{gather*}
we can choose $\varepsilon_0$ small enough so that for all $l>0$ there is no small denominator problem when solving equation~\eqref{eq:small denominator} for $W_1$.

Defining $H_2=U_1^*H U_1$ and repeating the above steps we arrive at the following:
\begin{lemma} \label{lemma:recursion}
We define $W_j$ inductively by $[\mathcal{D}_j,W_j]=\imath\mathcal{O}_j$, and furthermore $U_j=e^{\imath W_j}$, $H_{j+1}=U_j^*H_j U_j$.
Then we have:
\begin{plainlist}
\item[closeness to diagonal operator]
\begin{align}
\left\| \mathcal{D}_j(m,n;m,n) -
 \left[ B_c\left(\tfrac12+m\right)
   - \varepsilon_0  A^0_{m,m}   \right] \right \|
_{C^2(S^1)}  < \varepsilon_0\delta_j \label{eq:closeness}
\end{align}
\item[off-diagonal smallness]
\begin{subequations}
\begin{align}
\sum_{l\in\Z_+} \left\| \mathcal{O}_j(0,n;l,n)  \right\|_{C^2(S^1)} (l^s+1) &< \gamma_j   \\
\sum_{k\in\Z\setminus\{n\}}\sum_{l\in\Z_+} \left\| \mathcal{O}_j(0,n;l,k)  \right\|_{C^2(S^1)} e^{2\delta|n-k|} (l^s+1) &< \varepsilon_1\gamma_j \label{eq:offdiag smallness b}
\end{align}
\end{subequations}
\item[smallness of mixed terms]
\begin{subequations}
\begin{align}
\sum_{l\in\Z_+} \left\| \mathcal{M}_j(m,n;l,n)  \right\|_{C^2(S^1)} (l^s+1) &< (m^{s+1}+1)\delta_j   \\
\sum_{k\in\Z\setminus\{n\}}\sum_{l\in\Z_+} \left\| \mathcal{M}_j(m,n;l,k)  \right\|_{C^2(S^1)} e^{2\delta|n-k|} (l^s+1) &< (m^{s+1}+1) \varepsilon_1\delta_j
\end{align}
\end{subequations}
\end{plainlist}
Furthermore, we have the relations 
\begin{subequations}
\begin{align}
\gamma_{j+1}&=\const\gamma_j \delta_j, \\
\delta_{j+1}&=\delta_j(1+\const\gamma_j), \\
W_j(0,n;l,k) &\sim \frac{\mathcal{O}_j(0,n;l,k)}{l+1}\text{ as }l\to\infty, \label{eq:asymp}
\end{align}
\end{subequations}
where the constants depend on $s$.
\end{lemma}
Later we will see that 
\begin{align*}
\delta_j &< \const_s \varepsilon_0 \\
\gamma_j &< \left( \const_s \varepsilon_0 \right)^j
\end{align*}

\section{Estimates} \label{sec:estimates}
In order to prove Lemma~\ref{lemma:recursion}  we only need to check the induction hypothesis.
For $j=1$ we have the following non-vanishing matrix elements:
\begin{subequations}
\begin{align}
\mathcal{D}_1(m,n;m,n) &\stackrel{\phantom{n\ne k}}=  B_c\left(\tfrac12+m\right)
 + \tfrac12 \varepsilon_0^2  \widetilde{ A^0}_{m,m}\left({\beta(\xi+n)} \right) 
  - \varepsilon_0  A^0_{m,m} \left({\beta(\xi+n)} \right)  \notag \\
& \quad + 
  \tfrac12\varepsilon_1^2 \widetilde{ A^{1}}^{(0)}_{m,m}\left({\beta(\xi+n)} \right)  
  - \varepsilon_1  \widehat{ A^{1}}^{(0)}_{m,m}\left({\beta(\xi+n)} \right); \\
\mathcal{M}_1(m,n;l,k) &\stackrel{n\ne k}= 
  \tfrac12\varepsilon_1^2 \widetilde{ A^{1}}^{(n-k)}_{m,l}\left({\beta(\xi+n)} \right)  
  - \varepsilon_1  \widehat{ A^{1}}^{(n-k)}_{m,l}\left({\beta(\xi+n)} \right), \\
\mathcal{M}_1(m,n;l,n) &\stackrel{\phantom{n\ne k}}= 
  \tfrac12 \varepsilon_0^2  \widetilde{ A^0}_{m,l}\left({\beta(\xi+n)} \right) 
  - \varepsilon_0  A^0_{m,l} \left({\beta(\xi+n)} \right)  \notag \\
& \quad + 
  \tfrac12\varepsilon_1^2 \widetilde{ A^{1}}^{(0)}_{m,l}\left({\beta(\xi+n)} \right)  
  - \varepsilon_1  \widehat{ A^{1}}^{(0)}_{m,l}\left({\beta(\xi+n)} \right); \\
\mathcal{O}_1(m,n;0,n) &\stackrel{\phantom{n\ne k}}= 
  \tfrac12 \varepsilon_0^2  \widetilde{ A^0}_{m,0}\left({\beta(\xi+n)} \right) 
  - \varepsilon_0  A^0_{m,0} \left({\beta(\xi+n)} \right)  \notag \\
& \quad + 
  \tfrac12\varepsilon_1^2 \widetilde{ A^{1}}^{(0)}_{m,0}\left({\beta(\xi+n)} \right)  
  - \varepsilon_1  \widehat{ A^{1}}^{(0)}_{m,0}\left({\beta(\xi+n)} \right), \\
\mathcal{O}_1(m,n;0,k) &\stackrel{n\ne k}=
  \tfrac12\varepsilon_1^2 \widetilde{ A^{1}}^{(n-k)}_{m,0}\left({\beta(\xi+n)} \right)  
  - \varepsilon_1  \widehat{ A^{1}}^{(n-k)}_{m,0}\left({\beta(\xi+n)} \right) 
\end{align}
\end{subequations}
Note that we may assume $\widehat{ A^{1}}^{(0)}_{m,l}=0$; we can always achieve this by changing $A^1$ by a function of $y$ only, this does not change $B$.
$\widetilde{ A^{1}}^{(0)}_{m,l}=0$ cannot be made disappear in the same way because it involves the average of $\left(A^1\right)^2$ instead of $A^1$.

If we rewrite the induction hypotheses \eqref{eq:closeness}-\eqref{eq:asymp} in terms of the coefficients of the vector potential we get:
\begin{subequations}
\begin{multline}
\sum_{l\in\Z_+} \Big \| \tfrac12 \varepsilon_0  \widetilde{ A^0}_{m,l}\left({\beta(\xi+n)} \right) 
  -   A^0_{m,l} \left({\beta(\xi+n)} \right)   \\
  +  
  \tfrac12\varepsilon_1 \widetilde{ A^{1}}^{(0)}_{m,l}\left({\beta(\xi+n)} \right)  
  -  \widehat{ A^{1}}^{(0)}_{m,l}\left({\beta(\xi+n)} \right) \Big \|
 < \const \left(m^{s+1}+1\right) \label{eq:estimate diag} 
\end{multline}
\begin{multline}
\sum_{k\in\Z}\sum_{l\in\Z_+} \Big \| 
  \tfrac12\varepsilon_1 \widetilde{ A^{1}}^{(n-k)}_{m,l}\left({\beta(\xi+n)} \right)  
  -  \widehat{ A^{1}}^{(n-k)}_{m,l}\left({\beta(\xi+n)} \right) \Big \| e^{2\delta|n-k|} \\
 < \const \left(m^{s+1}+1\right) \label{eq:estimate offdiag}
\end{multline}
\end{subequations}
Since the vector potentials $A^0, A^1$ are periodic in $y$, equations \eqref{eq:estimate diag} and \eqref{eq:estimate offdiag} can be derived from the smoothness (analyticity) assumption and estimates for 
\[ \int_\R e^{\imath \eta y}\Omega_m(y)\Omega_l(y)\,dy \] in $\eta$.
For this we can use the estimates from the case of constant magnetic field \cite[lemmata 6-8]{DinSinSos:SLLLSPLMSPP}.
Because of our Remark~\ref{remark:weber hermite} there are no new estimates to prove.

\section{Finishing the Proof} \label{sec:proof}
The columns of the tranformation $U$ define new basis vectors $b_{\xi,n,m},n\in\Z,m\in\Z_+$.
From the induction hypothesis~\eqref{eq:offdiag smallness b} we get
\begin{align}
 |b_{\xi,n,m}|(m^{N+1}+1) e^{2\delta|n-j|} < \const. \label{eq:proofestimate}
\end{align}
This implies that for each $m$, we can apply the main results of \cite{Din:SPSTDOQC} to the operator $H_m$,
which is an ergodic family (in $\xi$) of difference operators with exponentially decaying almost periodic coefficients.
As a consequence we get part~\ref{eigenvalues} and the eigenfunction expansion in part~\ref{eigenfunctions}, equations \eqref{eq:maincoefficients} and \eqref{eq:maineigenfunctions}, of Theorem~\ref{theorem:main}.

Uniform convergence of \eqref{eq:maineigenfunctions} and its derivatives follows from \eqref{eq:proofestimate} and the relation \eqref{eq:Omega'} for the derivatives. As in \citep{DinSinSos:SLLLSPLMSPP}, \eqref{eq:proofestimate} also gives the estimate on the rate of decay in $y$, \eqref{eq:maindecay}.

Finally, from part~\ref{eigenvalues} we know that, for almost every $\xi$, 
the spectrum of $H_{m,\xi}$ is given by $\left(\lambda_m(\beta(\xi+n))\right)_{ n\in\Z}$.
If $E_{\xi,m,n}$ are the corresponding eigen\-spa\-ces, then $E_{m,n}:=\int^\oplus E_{\xi,m,n}\,d\xi$ are invariant subspaces such that the restriction of $H$ to $E_{m,n}$ is unitarily equivalent to multiplication with the function $\xi\mapsto\lambda_m(\beta(\xi+n))$.
On the other hand, $H_m=\bigoplus_{n\in\Z}E_{m,n}$.
Therefore, since $H_m=\int^\oplus H_{m,\xi}\,d\xi$, the operator $H_m$ is unitarily equivalent to multiplication with the function $\lambda_m$.

This together with the quantitive estimate~\eqref{eq:eigenvalues} finally gives the assertion in part~\ref{spectrum} about the measure of the spectrum.



\ifx\undefined\allcaps\def\allcaps#1{#1}\fi
  \ifx\undefined\nop\newcommand{\nop}[1]{}\fi
  \ifx\undefined\single\newcommand{\single}[1]{#1}\fi
  \ifx\undefined\SwapArgs\newcommand{\SwapArgs}[2]{#2#1}\fi
  \ifx\undefined\translationof\newcommand{\translationof}{English Translation
  of }\fi \ifx\undefined\submitted\newcommand{\submitted}{Submitted }\fi
  \ifx\undefined\submittedto\newcommand{\submittedto}{Submitted to }\fi
  \ifx\undefined\privcomm\newcommand{\privcomm}{Private communication}\fi
  \providecommand{\inpreparation}{In preparation}
  \providecommand{\toappearin}{To appear in }

\end{document}